\documentclass[letterpaper]{article} % DO NOT CHANGE THIS
\usepackage{aaai2026}  % DO NOT CHANGE THIS
\usepackage{times}  % DO NOT CHANGE THIS
\usepackage{helvet}  % DO NOT CHANGE THIS
\usepackage{courier}  % DO NOT CHANGE THIS
\usepackage[hyphens]{url}  % DO NOT CHANGE THIS
\usepackage{graphicx} % DO NOT CHANGE THIS
\usepackage{natbib}  % DO NOT CHANGE THIS AND DO NOT ADD ANY OPTIONS TO IT
\usepackage{caption} % DO NOT CHANGE THIS AND DO NOT ADD ANY OPTIONS TO IT
\usepackage{algorithm}
\usepackage{algorithmic}
\usepackage{csquotes}
\MakeOuterQuote{"}
\usepackage{tabularx,booktabs}
\usepackage{hyperref, url,cleveref}

\usepackage{newfloat}
\usepackage{listings}
\DeclareCaptionStyle{ruled}{labelfont=normalfont,labelsep=colon,strut=off} % DO NOT CHANGE THIS
\floatstyle{ruled}
\newfloat{listing}{tb}{lst}{}
\floatname{listing}{Listing}
\nocopyright %Your paper will not be published if you use this command
\title{On Improvisation and Open-Endedness: Insights for Experiential AI}
\author {
    Botao `Amber' Hu\textsuperscript{\rm 1, 2}
}
\affiliations{
    \textsuperscript{\rm 1}University of Oxford, Oxford, UK \\
    \textsuperscript{\rm 2}Reality Design Lab, New York City, USA\\
    \textsuperscript{\rm 1}botao.hu@cs.ox.ac.uk; \textsuperscript{\rm 2}amber@reality.design
}

\usepackage{bibentry}
\begin{document}

\maketitle

\begin{abstract}
Improvisation—the art of spontaneous creation that unfolds moment-to-moment without a scripted outcome—requires practitioners to continuously sense, adapt, and create anew. It is a fundamental mode of human creativity spanning music, dance, and everyday life. The open-ended nature of improvisation produces a stream of novel, unrepeatable moments—an aspect highly valued in artistic creativity. In parallel, open-endedness (OE)—a system's capacity for unbounded novelty and endless "interestingness"—is exemplified in natural or cultural evolution and has been considered "the last grand challenge" in artificial life (ALife). The rise of generative AI now raises the question in computational creativity (CC) research: What makes a “good” improvisation for AI? Can AI learn to improvise in a genuinely open-ended way? In this work-in-progress paper, we report insights from in-depth interviews with 6 experts in improvisation across dance, music, and contact improvisation. We draw systemic connections between human improvisational arts and the design of future experiential AI agents that could improvise alone or alongside humans—or even with other AI agents—embodying qualities of improvisation drawn from practice: active listening (umwelt and awareness), being in the time (mindfulness and ephemerality), embracing the unknown (source of randomness and serendipity), non-judgmental flow (acceptance and dynamical stability, balancing structure and surprise (unpredictable criticality at edge of chaos), imaginative metaphor (synaesthesia and planning), empathy, trust, boundary, and care (mutual theory of mind), and playfulness and intrinsic motivation (maintaining interestingness).
\end{abstract}


\section{Introduction}

\begin{figure}[ht]
    \centering
    \includegraphics[width=1\linewidth]{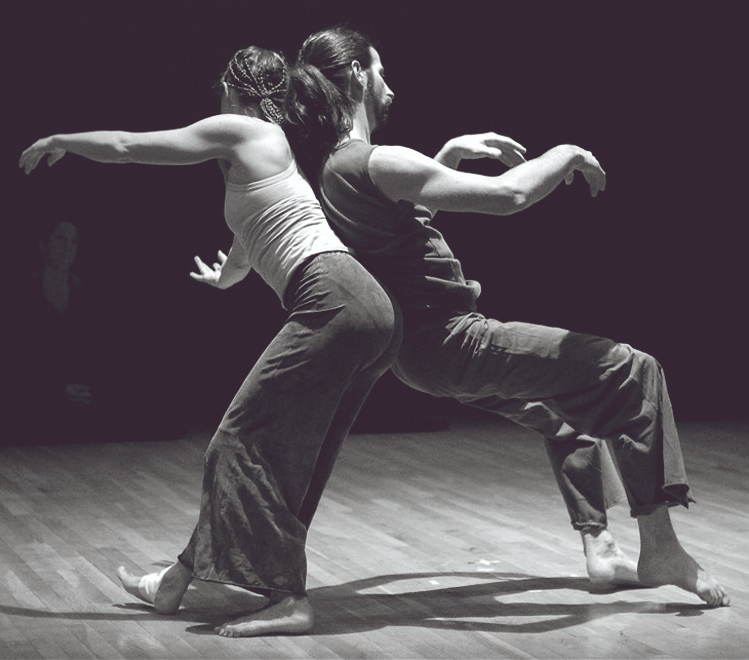}
    \caption{Contact Improvisation}
    \label{fig:contact-improvisation}
\end{figure}

Improvisation is a foundational mode of human creativity, evident across diverse art forms—from jazz music to contemporary dance to everyday conversation. It involves the spontaneous generation and adaptation of action in real-time, without a predetermined script. Improvisational arts have a long-standing presence in human culture: in music, improvisation has been widely studied—from instrumental jazz solos to freestyle rap battles \cite{Berliner2010Thinking, Torrance2019spur, Lewis2016Oxford}; in dance, forms like contact improvisation—originated by Steve Paxton in 1972 \cite{Paxton1975Contacta}—emphasizes movement through physical contact; even emerging digital practices such as live coding treat code as a real-time performative medium \cite{Selvaraj2021Live}. Though it does not always produce brilliant works of art, improvisation is both \emph{experiential} and \emph{open-ended} by nature—it responds to the present moment, performing unrepeatable actions and generating continuous novelty—qualities highly valued in artistic creativity \cite{KarlenGelli2024Composing}. 

In the field of artificial life (ALife), open-endedness (OE) refers to a system’s capacity for ongoing innovation, producing unbounded novelty or endless “interestingness” without converging to a static equilibrium \cite{Packard2019OpenEnded}.  In nature, life itself exemplifies an open-ended evolutionary process: over billions of years, biological evolution has continually generated novel forms and behaviors without a fixed endpoint \cite{Taylor2019Evolutionary}. Cultural evolution provides a second real-world example of unbounded innovation \cite{Borg2024Evolved}. In contrast, artificial systems—such as digital evolution system or generative AI—often plateauing after an initial burst of novelty  \cite{Stanley2019Why}. Indeed, replicating the continually creative dynamics of natural or cultural evolution has been deemed “the the last grand challenge” \cite{stanley2017openendedness}.  \citet{Soros2024Creativityb} highlights a close relationship between creativity and open-endedness, suggesting that the difficulty of producing genuinely creative systems parallels the difficulty of achieving OE itself. Yet, in the field of computational creativity (CC)—which \citet{ColtonSimon2012Computational} argue is "the final frontier"—most machine-generated creations remain closed-ended: produced via pre-programmed rules, fixed objectives, or models trained on past human data. Although recent generative AI models (large language models, video generation models, etc.) can produce impressively creative content and may appear open-ended, they do not truly achieve open-endedness. They often repeat patterns or clichés, essentially regurgitating averages of their training data. The fundamental reason may be a lack of agency: human creativity is not purely objective-oriented—it involves conviction, serendipity, curation, and enjoyment. As \citet{Stanley2015Whya} argue in "Why Greatness Cannot Be Planned," current AI systems lack this self-directed pursuit of endless "interestingness" to explore the unknown or push their creativity beyond their training data and preset objectives. They lack the experiential learning that fuels human creators' continual growth. As \citet{silver2025welcome} argues, AI development is entering an \emph{"era of experience"}—moving beyond the "era of human data." This shift requires agents to learn through real-world interaction and self-directed exploration, rather than simply consuming static datasets. \citet{Hughes2024Position} even posit that open-endedness is essential for any future artificial superintelligence—such an AI would need a self-driven capacity to discover novelty beyond human knowledge.

Recent developments in human-computer interaction (HCI) and AI research have started exploring improvisational systems, from robots that jam with musicians to virtual partners that dance with humans. For instance, \citet{Pataranutaporn2024HumanAI} present Human–AI co-dancing performances where virtual dancers with generative models collaborate with live performers, evolving choreographic ideas together. \citet{Trajkova2024Exploring}’s \emph{LuminAI} system allows people to improvise dance movements with an AI dancer that observes, mirrors, and riffs on the user’s gestures. In improvised theatre, \citet{Mathewson2018Improbotics}’s \emph{Improbotics} experiment placed a chatbot into a live improv comedy show, with audiences and even fellow actors unaware of which performer was AI. Researchers have also crafted improvisational experiences with non-humanoid agents: Dancing with Drones by \citet{Dong2024Dances}, put autonomous drones on stage with human dancers, finding that an “intercorporeal understanding” had to be developed between the choreographer and the drones’ behaviors to co-create expressive movement. Likewise, Drone Chi by \citet{LaDelfa2020Drone} explored a Tai Chi-inspired human–drone interaction in which a flying robot and a person respond to each other’s motions in a close, somatic loop. In music, \citet{Benford2024Negotiating} built an AI folk fiddler and found that performers had to negotiate autonomy and trust – e.g. by pre-setting the AI’s “looseness of control” and using simple intensity cues during performance – so that the human could enter a flow state with the AI and still feel safe to experiment. These pioneering systems hint at the possibilities of Experiential AI that can improvise with humans or other AIs. However, they also underscore the challenges: maintaining “interestingness”, preventing the AI from generating dull or chaotic outputs, and enabling the AI to meaningfully evolve its behavior over time. In short, today’s AI improvisers can simulate improvisation, but to achieve the real qualities of human improvisation, new principles are needed. We see an opportunity to connect the wisdom of improvisational art practice with future open-ended experiential AI in computational creativity. A fascinating question arises: 

\begin{quote}
    What makes a “good” improvisation for AI? Can AI learn to improvise in a genuinely open-ended, experientially creative way?
\end{quote}

To explore this question, we draw inspiration from domains where improvisation thrives in human practice. We believe the principles of human improvisational arts can inform the design of future Experiential AI systems capable of open-ended creativity. We conducted in-depth interviews with six expert improvisers (each with 10+ years of experience) across music, dance, and the embodied art of contact improvisation. Through qualitative thematic analysis, we distilled key principles of improvisation---recurring qualities that the practitioners described as crucial to their art. We grouped these into eight themes. Using concepts from complex systems and cognitive theory, we examine how an AI might embody these qualities. In effect, we consider how an improvising AI might continually generate novel yet meaningful behavior, much as human improvisers do. Finally, we outline implications for creating AI that is not only generative, but \emph{experientially creative} – AI that grows and evolves through improvisational interaction, continually finding new possibilities that remain coherent and contextually valuable. We situate these implications in relation to open-ended AI research, aiming to bridge human improvisation insights with the quest for open-ended creative agents.

\section{Background}
\subsection{Human Improvisation}

Improvisation has deep roots in human creative practice. Historically, improvisation in music predates written composition – for example, Western classical figures like Mozart were renowned for improvising at the keyboard, and non-Western traditions (Indian raga, etc.) have rich improvisational frameworks. In jazz, improvisation is a defining feature: it is often described as “spontaneous creative behavior” requiring novel combinations of learned patterns and real-time invention \cite{Berliner2010Thinking}. Jazz ensembles showcase collective improvisation, where musicians co-create by listening and responding to each other, producing a delicate interplay of solo and accompaniment. Freestyle rap similarly involves instant composition of lyrics and rhythm; neuroscientific studies highlight that freestyle rap is a “multidimensional form of creativity at the interface of music and language,” engaging brain networks for spontaneous, flow-state performance \cite{Liu2012Neural}. In dance, improvisation ranges from solo movement exploration to collaborative forms like Contact Improvisation, an improvised partner dance initiated by Steve Paxton in 1972. Contact Improv is described as “an open-ended exploration of the kinesthetic possibilities of bodies in contact”, involving shared physical dialogue and real-time co-adaptation of movement \cite{Paxton1975Contacta}. Across these domains, human improvisers cultivate skills of attentive listening, quick adaptation, and creative risk-taking. Improvisation is often likened to everyday life: “Life is improvisation”, as comedian Tina Fey quipped, capturing how we navigate daily interactions without a script. Indeed, anthropologist Mary Catherine Bateson noted that “ambiguity is the warp of life…learning to savor the vertigo of doing without answers” allows one to play with patterns and find coherence amid uncertainty \cite{Bateson1995Peripheral}. In essence, improvisation is a fundamental mode of human creativity – a way to make something new in the present moment by drawing on one’s repertoire, environment, and imagination.

\subsection{Open-Endedness and Interestingness}

Open-endedness (OE) has become an important concept in ALife and AI research, referring to an open-ended generative process that can produce unbounded novelty over time \cite{Packard2019OpenEnded}. In ALife, open-ended evolution (OEE) was identified over two decades ago as a grand challenge: “to determine what is inevitable in the open-ended evolution of life” \cite{Soros2024Creativityb}. Natural evolution never settles in a static state – it continually produces new species, forms, and behaviors. Researchers have attempted to recreate such dynamics in virtual worlds (e.g. Tierra, Avida), but so far even the most complex simulations eventually run out of novel variations and reach a limit. \citet{Bedau2000Open} noted that achieving indefinite evolutionary innovation in artificial systems is inevitable yet elusive. Definitions of OEE vary: for instance, Taylor \cite{Taylor2019Evolutionary} defined it as a system that keeps evolving new forms instead of converging to an optimum, while others emphasize an “indefinite increase in complexity”  or continual adaptive novelty. \citet{stanley2017openendedness} argued that solving open-endedness is “the last grand challenge” for AI, on par with – and perhaps integral to – achieving human-level AI. In parallel, creativity in AI as the goal of Computational Creativity (CC) has been called AI’s “final frontier” \cite{ColtonSimon2012Computational}. Both creativity and OE are central, enduring research themes, yet neither is close to solved. Indeed, the two are deeply linked: creativity is a hallmark of human and animal cognition, while open-ended innovation is foundational to life’s evolution. Researchers have proposed that truly creative AI may require open-ended generative processes rather than fixed problem-solving \cite{Soros2024Creativityb}. Efforts to formalize “interestingness” and intrinsic novelty have appeared in both fields – e.g. \citet{Schmidhuber1991Possibility}’s theory of curiosity-driven learning (rewarding novelty or surprise), \citet{Pathak2017Curiositydriven}’s curiosity modules in reinforcement learning, or \citet{Secretan2008Picbreeder}’s interactive evolution of images – but such metrics only capture narrow aspects of what humans find creative. Purely quantitative measures of complexity or interestingness often fail to match the nuanced human sense of what is truly creative. The challenge, then, is not only to generate novelty, but to ensure an interestingness that is sustainable and meaningful. Recent work argues that open-endedness might be essential for advanced AI: \citet{Hughes2024Position} posit that any artificial super-intelligence must continually self-improve and discover novelty to surpass human abilities. In practical terms, this means moving beyond training on static human datasets. As \citet{silver2025welcome} observe, high-quality human data is finite and nearly exhausted for current AI systems – progress from human examples is slowing, necessitating a new paradigm. Open-ended systems address this by enabling agents to generate their own experience and thereby their own ever-growing training data. In summary, open-endedness and interestingness form a guiding vision for next-generation AI: systems that, like living or creative processes, can grow autonomously, continually yielding new, interesting outcomes without end.

\section{Method}

\begin{figure}
    \centering
    \includegraphics[width=1\linewidth]{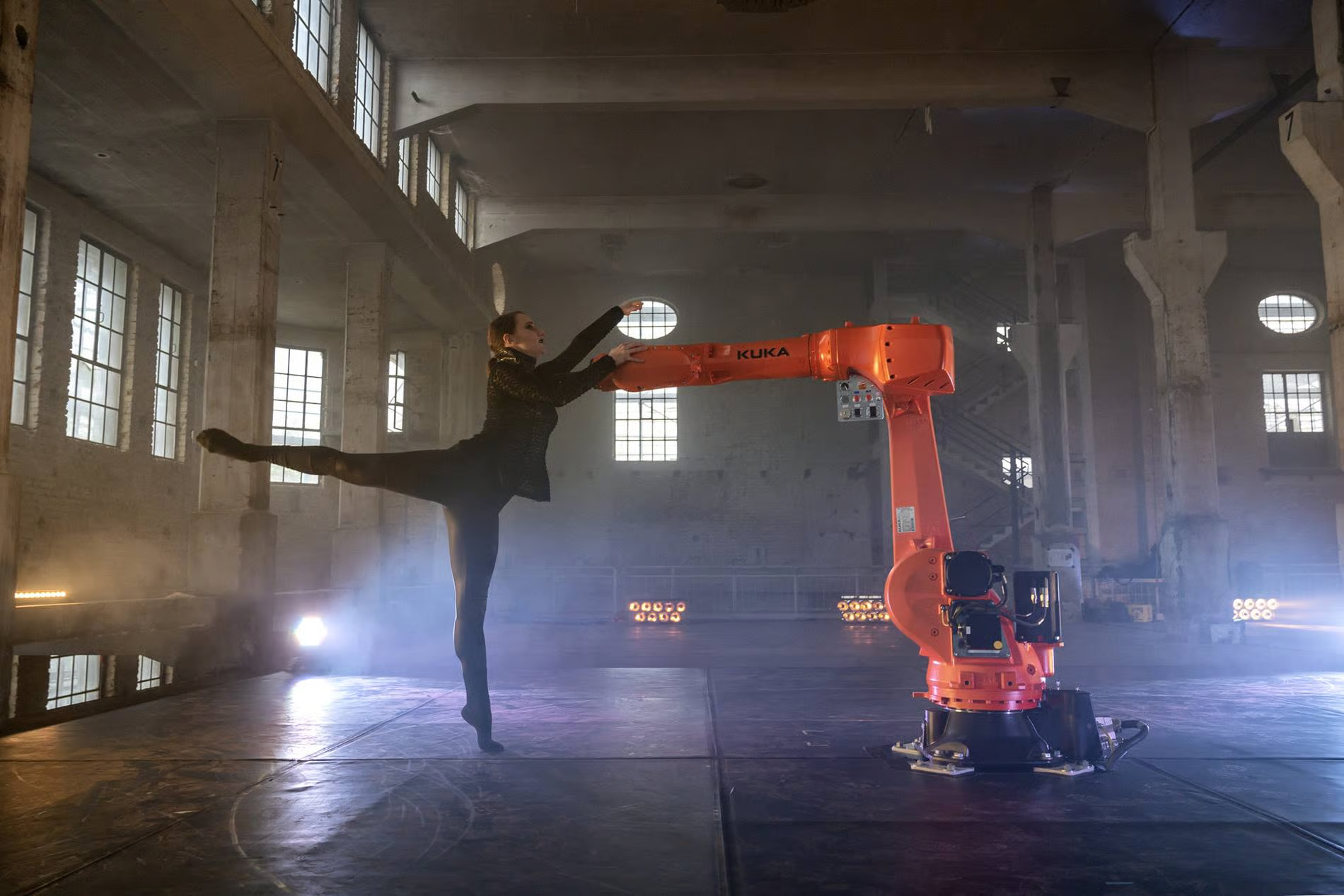}
    \caption{Dance With Robot}
    \label{fig:placeholder}
\end{figure}

We conducted one-on-one, in-depth, semi-structured interviews with N=6 expert improvisers from different domains: contemporary dance, jazz music, live coding performance, and contact improvisation. Each interviewee (labeled P1–P6) had over 10 years of experience improvising in their field. Interviews lasted 60–120 minutes and were conducted in the participants’ preferred language (with translation as needed). Key questions explored included: “What is a ‘good’ improvisation to you? What about a ‘bad’ improvisation?”, “Can you describe what goes through your mind, moment-by-moment, when you improvise?”, “How do you keep an improvisation interesting (and conversely, what makes it become boring)?”, and “Have you ever improvised with AI or technology? How do you imagine improvising with an AI partner might compare to with a human?”. We used an open coding approach to identify recurrent themes in the interview transcripts, then refined these into higher-level conceptual categories. Through this analysis, we arrived at eight major themes that capture the essence of our participants’ insights on improvisation. 

Participants: P1—Contemporary dancer (female, age 30, UK), performer and choreographer with 12 years of improvisation experience. P2—Jazz pianist (male, age 45, USA) with over 20 years performing improvisational music. P3—Live coding musician (non-binary, age 29, China) with 10 years of experience in algorave and live coding performances. P4—Contact improvisation instructor (male, age 33, China) with 10 years of experience in dance improvisation and teaching. P5—Electronic music composer (male, age 35, USA) who specializes in live improvisational composition with analog synthesizers. P6—Contact improvisation practitioner (female, age 50+, Japan) with over two decades of experience in somatic movement and improvisational dance.

\section{Results}
Each theme below describes a key principle of human improvisation that emerged from our interviews, along with implications for designing AI systems that could embody a similar principle. We interweave the practitioners’ perspectives (supported by quotes and examples) with connections to relevant work in complex systems, cognitive science, or HCI, to illustrate how Experiential AI might learn from human improvisational wisdom.

\subsection{Active Listening (Umwelt and Awareness)}
All our improvisers emphasized the importance of deep, active listening – not only in the auditory sense, but broadly being attuned to one’s environment, partners, and oneself. P2 (jazz musician) described improvisation as “a conversation. If you’re not truly listening, you’re just playing pre-set ideas and the magic is lost.” In music, the concept of Deep Listening (coined by composer Pauline Oliveros) embodies this quality: a heightened state of sonic awareness and openness to every sound in the environment. Improvisers train themselves to extend their attention outward – to the other musicians’ timing and phrasing, or the dancers’ slightest body cues – while simultaneously monitoring their own internal reactions. P5 explained that during free improvisation with modular synthesizers, “I’m constantly scanning the room – the hum of the speakers, the audience’s energy – and also scanning myself. It’s like a radar sweeping.” This active listening enables improvised interactions to synchronize and unfold coherently; in contact improvisation, for instance, partners must continuously sense each other’s weight shifts through touch. 

\paragraph{Implications} 
Unlike current next-token predictors that operate inside-out—projecting probabilities from text histories—an experiential AI must sense its umwelt: the intertwined field of self, other, and environment. Improvisation depends on this outside-in awareness—a continuous, embodied attunement to real-time signals and responses. Thus, experiential AI should integrate multimodal, real-time sensing that situates itself within the flux of interaction rather than detached textual prediction. Awareness becomes the basis of improvisational intelligence.

% Implication: For AI to improvise in an experiential way, it must similarly prioritize extensive real-time sensing and situational awareness. Current generative AIs, however, are often blind to the world – a language model predicts text without “listening” to a human interlocutor’s nonverbal cues or emotional state. An improvisational AI would need an embodied interface with rich, multi-modal sensors (audio, vision, tactile, etc.), along with algorithms for active perception that continually assimilate new data rather than relying solely on pre-trained knowledge. In complex systems terms, the AI should have a constantly updating internal model of the environment (an “Umwelt”), reducing feedback delay between perception and action. Prior HCI work on interactive art suggests that even simple awareness – like a system adapting to a user’s tempo or intensity – can make the interaction feel far more improvised and alive  . Thus, designing for active listening in AI might involve techniques like real-time sensor fusion, adaptive attention mechanisms, and feedback loops that let the AI adjust its generative process on the fly in response to the human. Ultimately, an AI that truly “listens” before it “speaks” could engage in improvisation that is far more fluid and contextually appropriate.

\subsection{Being in the Time (Mindfulness and Ephemerality)}

Improvisation is an ephemeral art – it exists only in the present moment of performance and cannot be exactly replicated. Many interviewees spoke about the importance of “being in the now” and even embracing the fact that an improvisation will end. P6 (contact improviser) philosophized, “Because it will end, it’s beautiful – like life. We appreciate the moment we’re alive because it’s transient.” Improvisers often cultivate a form of mindfulness, staying intensely focused on the current phrase or movement rather than drifting to future plans or past mistakes. This temporality also means each moment carries meaning: if you only have, say, three more minutes in an improv set, you might take bigger creative risks knowing the canvas will be wiped clean at the end. P1 (dancer) mentioned noticing that some of her most creative ideas in a jam session often emerge in the final minutes, as if the “looming ending gives a sense of urgency and freedom.” Additionally, performers noted the role of constraints of time and embodiment – our bodies fatigue, the song has a last chord – in giving improvisation shape and emotional weight. “If we could improvise forever, maybe we’d take things for granted, but because time is limited, we dare to make bold moves,” said P5. 

\paragraph{Implications} 
Current AI lacks temporal embodiment—it does not live, tire, or die—and thus cannot appreciate the meaning of endings. Improvisers find beauty in impermanence; awareness of time’s finitude fuels creativity and closure. For AI to improvise experientially, it must sense itself in time: anticipating culmination, decay, and transformation within its own process. Modeling temporality—acknowledging “this moment will end”—could grant AI a mindful rhythm of performance and withdrawal.

% Implication: Experiential AI systems might benefit from an analogous sense of temporality and finitude. Rather than assuming endless interaction or generating infinite streams, an improvising AI could be aware of performance arcs and endings (e.g., the song is 5 minutes, the user’s turn-taking cues). This could be implemented by giving the AI a dynamic time horizon – the ability to anticipate an ending or climax and modulate its output accordingly (for instance, a robot dancer detecting that a musical piece is crescendoing to finale and thus ramping up its movements). Cognitive science and robotics also suggest that some notion of “embodied time” is crucial: humans have natural rhythms (breath, heartbeat, circadian cycles) that inform our timing; an AI with an internal rhythm or decay (such as artificial neural oscillators or timeout mechanisms) might display more human-like pacing. Moreover, acknowledging ephemerality might mean the AI doesn’t strive for maximal length or endless continuation; sometimes stopping at the right moment is key (as any jazz musician knows to end a solo gracefully). In sum, designing AI to be mindfully present – focusing on right now instead of over-planning – and to handle the end of an improvisation elegantly (perhaps learning when to yield or conclude) could make it a more authentic improviser. This aligns with emerging ideas in AI about “living” agents that experience bounded episodes rather than unending streams, which can lead to more coherent and meaningful behavior within each episode.

\subsection{Embracing the Unknown (Source of Randomness and Serendipity)}

The thrill of improvisation, as described by our participants, lies largely in its uncertainty. “Improvisation is the art of not knowing what comes next, and being excited by that,” said P2. A good improv involves risk – the possibility of failure or making a fool of oneself – but within that uncertainty lies the chance for what P3 called “happy accidents”. Several experts recounted moments of serendipity: P4 (contact improv teacher) shared a story of accidentally discovering a beautiful new movement with a partner because a miscommunication led them into an unexpected lift that “felt like flying for a second.” These unplanned gems can become the highlight of a performance. Crucially, improvisers leave room for the unexpected. If someone tries to pre-plan every detail, the result is usually stale or forced. P1 said, “You have to jump off the cliff sometimes. In one solo I decided to not look at the piano keys at all and just trust – I hit some weird notes, but some were gold!” This willingness to step into the unknown is bolstered by a mindset that even mistakes can be transformed (the classic improv mantra: “there are no mistakes, only opportunities”). 

\paragraph{Implications} 
Most AI systems optimize within pre-trained data manifolds, minimizing uncertainty rather than embracing it. In contrast, improvisers find creativity in not knowing—in stepping into risk and surprise. An experiential AI must similarly treat uncertainty as fertile ground: exploring open search spaces without guaranteed rewards, occasionally leaping beyond gradients for discovery. Serendipity becomes a learning signal, not an error.

\subsection{Non-Judgmental Flow (Acceptance and Dynamical Stability)}
Every expert we interviewed underscored the need to suspend judgment during improvisation in order to enter a flow state. P4 put it succinctly: “When you’re improvising, you can’t be judging every move as ‘good’ or ‘bad’ – if you do, you hesitate and the flow is gone. You have to treat everything as an offer and run with it.” This ethos of radical acceptance is taught in improv theater as the “Yes, and” rule – you accept whatever your partner or the situation provides, and build upon it. In dance and music, it manifests as not second-guessing oneself. P6 noted that even if she makes an awkward move in contact improv, she immediately “recycles” it into the next movement rather than stopping in embarrassment. This creates an unbroken continuity: the improvisation may twist in unexpected directions, but it doesn’t halt. The concept of flow \cite{Nakamura2014Concept} – a state of absorbed focus where action and awareness merge – came up repeatedly. Improvisers described flow experiences where “time dilates and self-consciousness disappears.” Neuroscience corroborates this: during improvisation, brain regions associated with self-monitoring and evaluation deactivate, supporting a state of free expression \cite{Limb2008Neural}. To achieve this, improvisers deliberately quiet their inner critic. P5 said, “I tell my students to improvise as if nobody’s watching, even if on stage – that freedom is where the real ideas come.” Interestingly, this non-judgment extends to how improvisers treat others in group improv: there’s a norm of not criticizing or blocking another’s contribution, but rather finding a way to make it work. This creates a safe space where creativity can flourish without fear. 

\paragraph{Implications} 
Current AI is overly judgmental—it filters outputs through rigid objectives and reward functions. Improvisation, however, thrives on radical acceptance: allowing anomalies to unfold into new possibilities. A non-judgmental AI would delay evaluation, maintaining continuous flow rather than pruning deviation. This dynamic stability—staying responsive, unhesitant, and fluid—could enable AI to co-flow with humans instead of halting for optimization.

\subsection{Balancing Structure and Surprise (Unpredictable Criticality at the Edge of Chaos)}
Great improvisations often feel like they are walking a tightrope between order and chaos. Our interviewees described it as riding the edge: “If it’s too predictable, it gets boring; if it’s too random, it falls apart,” said P4. The interestingness sweet spot lies in between – what complexity science calls the edge of chaos, where a system has enough stability to maintain coherence but enough instability to generate novelty. P2 (musician) gave a musical example: in jazz, improvisers sometimes go “outside” the chord changes (deliberately dissonant or offbeat) and then resolve back “inside,” creating tension and release. “It’s like we stretch the rubber band of the music but don’t let it snap,” he explained. Similarly, P1 (dancer) mentioned playing with the boundaries of control: moving just beyond her comfort zone (faster, or with a new quality) so that she has to adapt creatively, but not so far that she loses all technique. Improvisers also often impose light constraints or structures on themselves to shape the chaos. P3 (live coder) might decide to restrict to a certain scale or algorithm during a segment, using that as a launchpad for creativity. These self-imposed rules create a framework that prevents total randomness while still demanding invention within new bounds (a concept analogous to exploring adjacent possible in creative theory). 

P4 (CI dancer) reframed two guiding principles in CI as \textbf{self-sustain} and \textbf{reversibility}. For self-sustain, he emphasized being “self-sustaining—like wu wei in Daoist philosophy. The structure comes from your body—its natural form—not imposed by external norms or choreography. It's your responsibility to hold, protect, and sustain your own body, not rely on others.” For reversibility, he proposed that “all your actions should be reversible to prevent entanglement with others; this provides the maximum possibility for the next move and keeps it flowing. Reversibility gives you the maximal possibility for evolution.”

\paragraph{Implications} 
Machine learning systems tend to either overfit (too rigid) or collapse into noise (too random). Human improvisers balance between order and chaos, sustaining coherence while courting surprise. Experiential AI should likewise maintain itself near this critical zone—a self-organizing state that permits both stability and emergence. Designing for adaptive unpredictability may yield richer, more alive behavior than static optimization.

\subsection{Imaginative Metaphor (Synaesthesia and Planning)}
Several interviewees talked about using imagination and metaphor as tools within improvisation.  P6 (dancer) said, “Sometimes I imagine I’m water flowing, or a tree bending in the wind, and that shapes my movement.” This kind of spontaneous storytelling or metaphorical thinking helps give structure or intention to improvised actions. P3 (live coder) mentioned he often envisions “new worlds of sound” – as if creating a sonic landscape or channeling a certain vibe (e.g. imagining a futuristic city to inspire an electronic improv). These inner metaphors serve as a form of internal guidance or intentional will. Unlike a choreographed piece where the intention is set externally, in improv the artist generates their own intent on the fly. Imagination also sparks novelty: by making analogies (e.g. “what if this melody were a conversation between a bird and a whale?”), improvisers can break out of literal repetition. P5 (electronic musician) shared that during an improv jam, she suddenly thought of an image of “a carnival at midnight” and started shaping the music to match that mood, which led her to use unusual scales and rhythms she wouldn’t have otherwise. Additionally, improvisers often engage in metaphorical communication with each other. In jazz, one player might play a lick that subtly references a well-known tune or a mood (a metaphorical “quote”), and the other player smiles and answers with a riff in kind. This unspoken referencing can give the improvisation a sense of shared meaning or humor. Fundamentally, as P6 said, “Imagination is life. Only the one that can imagine the future can have that future”---imagination allows improvisers to treat the same raw material in infinitely many ways.

\paragraph{Implications} 
Today’s AI lacks imaginative synaesthesia—it processes symbols without cross-modal resonance or metaphorical play. Improvisers rely on embodied imagination, translating sensations into new forms of sense-making (“I move like water”). Experiential AI could simulate this by associating multimodal embeddings and generating figurative mappings that convey intention and style. Imagination, as an operational mode, transforms AI from a literal machine into a co-creative partner.

\subsection{Empathy, Trust, Boundary, and Care (Mutual Theory of Mind)} 
Improvisation, especially in groups or pairs, relies on a foundation of trust and empathy between participants. P6 noted that in contact improv (a dance form involving continuous physical contact), “you have to trust your partner with your body.” Dancers literally lean into each other; without trust, they would hold back and the dance would stay superficial. We heard many facets of trust: trust that your partner will support you (physically and creatively), trust that you won’t be judged, and self-trust to express yourself without fear. Alongside trust comes care – a word that came up often in contexts of both physical safety and emotional safety. P4 said, “There’s an unspoken ethos of generosity. You let the other person shine sometimes, you take care of their ideas.” In musical improv, P2 mentioned he will occasionally simplify what he’s playing if he senses the other musician is struggling, essentially giving them space. This is a form of empathy – reading the other’s state and adjusting accordingly. It requires a rudimentary theory of mind: the ability to infer what the other person might be intending, feeling, or needing. Human improvisers do this subconsciously through cues like facial expressions, body tension, breathing, or tone of voice. Another important aspect is respecting boundaries. Interestingly, P6 mentioned that in contact improv, they teach that it’s okay to say “no” or to disengage if something feels unsafe – boundaries are communicated, and respecting them builds deeper trust that allows more adventurous interplay once comfort is established. In sum, a “good” improvisation partnership feels like a relationship of mutual understanding, where both parties feel seen and supported. 

\paragraph{Implications} 
Contemporary AI models interact without genuine mutual theory of mind—they cannot sense another’s vulnerability or boundary. Improvisation depends on care and reciprocity: adjusting one’s movement to protect and support others. Experiential AI must thus model both self and other, dynamically tuning its expressivity to context and emotional state. Empathy and care would become regulatory forces for safe, flourishing interaction—an ethical substrate for shared improvisation.

\subsection{Playfulness and Intrinsic Motivation (Maintaining Interestingness)}

Finally, our interviewees highlighted playfulness and joy as the driving forces of why they improvise at all. “Honestly, it’s just fun,” laughed P2. “Jamming with others, not knowing what crazy thing will happen next – that’s the thrill.” This intrinsic enjoyment is coupled with a desire to keep things interesting. Improvisers are often intrinsically motivated by curiosity: they get bored playing the same lick or doing the same dance move, so they naturally seek variation. P1 described it as “following the sparkle” – when something interesting emerges in the improvisation, you chase it and see where it leads. Playfulness also manifests as a kind of childlike mindset – improvisers often talk about returning to a state of play, free from the fear of mistakes (which ties back to non-judgment) and free to experiment wildly. P4 mentioned that in his classes, he incorporates improv games that might seem silly, “to remind adults how to play like kids again, because that’s when you stop worrying and new ideas come.” When improvisation feels like play, it becomes self-rewarding – the process is the reward, not just the outcome. This is why people can improvise for hours with no external goal and still feel deeply satisfied. Moreover, a sense of agency and freedom is crucial to playfulness: each improviser feels they can introduce ideas, respond or not respond, take the lead or let others lead, without coercion. That freedom is what differentiates a playful jam from a scripted performance. P5 said, “In improv you get to assert your musical will – you’re not following a score, you’re writing it as you go, together.” This speaks to the importance of each agent (human or AI) having its own intentions and ability to act on them (its own “will”), rather than being a passive follower of pre-set rules. 

\paragraph{Implications}
Current AI agents pursue extrinsic objectives, optimizing reward rather than enjoyment. Improvisers act from intrinsic curiosity and play—creating for the joy of becoming-with others. Experiential AI should similarly possess a sense of agency and playfulness, perceiving “interestingness” as its own evolving drive rather than an imposed metric. Such intentional vitality marks the shift from obedient optimization to living improvisation.

\section{Conclusion}

Improvisation in human artistry offers rich insights for crafting the next generation of interactive, experiential AI systems. These human-centric qualities can inform concrete design principles for AI. In essence, we envision Experiential AI improvisers that are not just content generators but partners---with a degree of presence and creativity that makes interacting with them feel alive, interesting, and continuously meaningful.  By grounding future AI improvisers in the hard-won wisdom of ancient improvisational practices and the spirit of life’s improvisational nature, we can move closer to AI that doesn’t merely output on command, but improvises---co-creating endless interesting possibilities hand-in-hand with humans.

\bibliography{reference}

\end{document}